\documentclass[a4paper]{jpconf}
\usepackage{graphicx}
\usepackage{amsmath}
\usepackage{amsfonts}
\usepackage{amssymb}
\usepackage{amsthm}
\usepackage{caption}
\usepackage{fontenc}
\usepackage{graphicx}
\usepackage{comment}
\usepackage{wrapfig}
\usepackage{epsfig}
\newcommand\ddfrac[2]{\frac{\displaystyle #1}{\displaystyle #2}}
\newcommand{\si}{\sigma_{eff}}
\usepackage[most]{tcolorbox}
\usepackage{textpos}
\usepackage{longfbox}
\begin{document}
\title{The proton structure via double parton scattering}

\author{Matteo Rinaldi}

\address{Dipartimento di Fisica e Geologia. Universit\`a degli studi di 
Perugia. Istituto Nazionale di Fisica Nucleare, section of Perugia. 
Perugia via A. Pascoli, I-06123, Italy.}

\ead{matteo.rinaldi@pg.infn.it}

\begin{abstract}
In this talk
we present the  results of the investigation on the so called
double parton distribution functions (dPDFs),  accessible quantities in 
high
energy proton-proton and proton nucleus collisions, in double parton 
scattering 
processes (DPS). These new and almost unknown distributions encode  
information on how partons inside a proton are 
correlated among each
other and  represent a new tool to explore the three dimensional 
partonic 
structure of hadrons. 
In the present contribution, results of the calculations of
dPFDs are  presented also
including phenomenological investigations
on the impact of double correlations in
 experimental observables, 
  showing how the latter
 could be observed in the next LHC run. In addition we 
discuss how present information on experimental observables could be 
related to the transverse proton structure.

\end{abstract}

\section{Introduction}
Thanks to the high luminosity reached in collider experiments, such as 
at the LHC, the investigation on multiple parton interactions (MPI) 
became  relevant in the study of  
 hadron-hadron collisions. 
In these kind of events, more than one parton of a hadron 
interact with  partons of the other colliding hadron. However, the
 MPI contribution  to the total cross section is  
suppressed  with respect to the single partonic interaction.
Nevertheless,
MPI  represent a background for the search of new  Physic at the 
LHC and 
the measurement of their cross sections  is an 
important experimental
 challenge. From a theoretical point of view,  one of the main 
interests on MPI is
the 
possibility of accessing new fundamental information on the partonic 
proton 
structure \cite{hadronic,noij3}. To this aim,  we focused our attention 
on the   double 
parton 
scattering (DPS), the most simple case of MPI, which can be 
observed, in principle, 
 in several processes, e.g., $WW$  with dilepton productions 
and double
Drell-Yan processes (see, Refs. \cite{3a,4a,5a,6a, report} for
recent 
reviews).
At the LHC, DPS has been
observed some years ago \cite{16a} and represents  also a
background for the Higgs production in several channels.
Formally, 
the DPS cross section
depends on  the so called double 
parton distribution functions (dPDFs),
$F_{ij}(x_1,x_2,{\vec z}_\perp,\mu)$, which describe the 
joint probability 
of finding two partons of flavors $i,j=q, \bar q,g$ with 
longitudinal momentum fractions $x_1,x_2$ and distance 
$\vec z_\perp$ in the transverse plane inside the hadron, see Ref. 
\cite{1a}~.
Here $\mu$ is the renormalization scale.
Since for the moment being no data are available for dPDFs, they 
are 
 usually approximated in experimental analyses through a fully 
factorized 
ansantz:

\begin{align}
 F_{ij}(x_1,x_2, \vec z_\perp,\mu) = F_i(x_1,\mu)F_j(x_2,\mu) T(\vec 
z_\perp),
 \label{app}
\end{align}
being $F_i(x,\mu)$ the usual
 one-body parton distribution functions (PDFs) and  $T(\vec z_\perp)$, 
a 
distribution 
encoding the probability of 
finding two partons at distance $z_\perp$.
In  assumption  (\ref{app}) all possible unknown  
 double parton
correlations (DPC) between the two interacting partons  (e.g.  Ref. 
\cite{kase1}) have been neglected.
However, 
 dPDFs  are non
perturbative quantities in QCD and  they can not be easily evaluated, 
thus
constituent quark model (CQM) calculations   could help to grasp their 
basic 
features.
In particular we studied
 to which extent  the factorized approximation (\ref{app}) is a 
suitable 
ansatz for dPDFs see Refs. \cite{noij3,man,noi1,noij1,bru, 
noice}~. 
Double PDFs
  are first of all calculated at the low
hadronic scale of the model, $\mu_0$, then, in order
 to compare the outcome with future data taken at
high momentum transfer, $Q > \mu_0$, it is  necessary to perform a
perturbative QCD (pQCD) evolution  by 
using  dPDF evolution equations, see Refs. \cite{23a,24a,blok_1}.
In particular, this step is fundamental to understand to what extent  
DPC 
survive at the kinematic conditions of experiments.
Thanks to this procedure, future data analyses
of the DPS processes could be guided, in principle, by 
model calculations, see  Ref. \cite{noiprl}.
To this aim in Refs. \cite{noij1,noij2}~, DPC in dPDFs have been 
studied at  
the energy 
scale of the experiments and  the so called  
$\sigma_{eff}$,  evaluated in Refs. \cite{noij3,noice,noiprl,noiPLB, 
noiads}~. In fact,  DPS cross 
section, in 
processes with final 
state 
$A+B$, is written through   the following ratio  (see 
e.g.~Ref.~\cite{MPI15}):

\vskip -0.5cm
\begin{eqnarray}
\sigma^{A+B}_{DPS}  = \dfrac{m}{2} \dfrac{\sigma_{SPS}^A 
\sigma_{SPS}^B}{\sigma_{eff}}\,,
\label{sigma_eff_exp}
\end{eqnarray}

where $m$ is combinatorial factor depending on the final states $A$ and 
$B$ 
($m=1$ for $A=B$ or $m=2$ for $A \neq B$) and
$\sigma^{A(B)}_{SPS}$ is the single parton scattering cross section with 
final 
state $A(B)$. 
The present knowledge on DPS cross sections
has been condensed in the experimental  extraction of  ~
$\sigma_{eff}$~
\cite{MPI15,S1,S2,S3,S4,S5,S6,S7,data8,data9,data6,data10,data11,data12}
. 
A constant value, $\sigma_{eff} \simeq$ 
15 mb, is compatible, within errors, with data: 
a
result  obtained within  the fully uncorrelated ansatz for 
dPDFs   (all DPC are neglected). In the next sections the  results of 
the calculations
$\sigma_{eff}$   within CQM will be described in
order to characterize 
``signals'' of DPC.

\section{Constituent Quark Models Calculations of dPDFs}
\begin{figure}[t]
\includegraphics{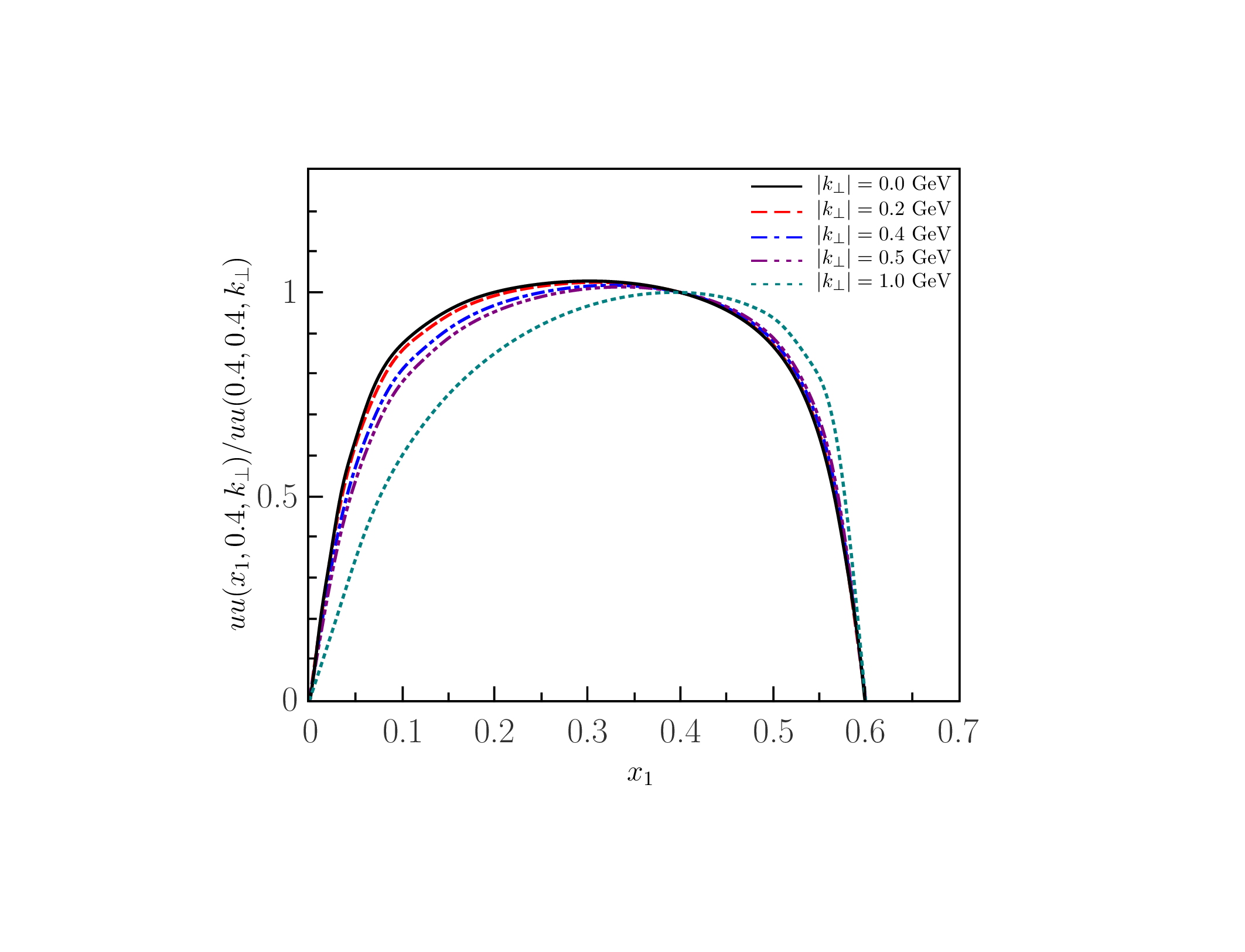}
\includegraphics{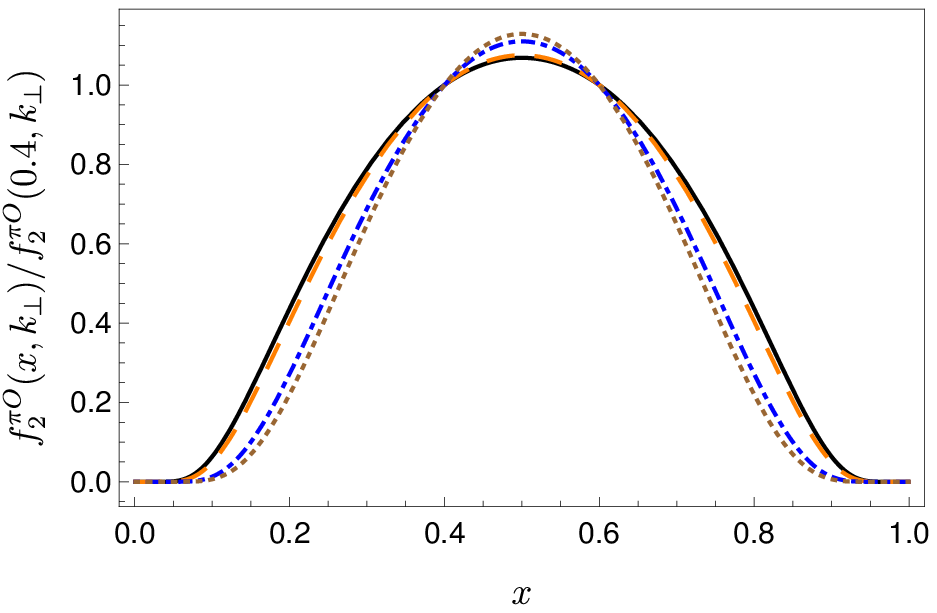}
\vskip 2cm
\caption{\footnotesize{ Left panel: the ratio 
$uu(x_1,0.4,k_\perp)/uu(0.4,0.4,_\perp)$ evaluated within a 
relativistic quark model, see Ref. \cite{noij1} as a function of $x_1$ 
and different values of $k_\perp$. Right panel: same ratio of the left 
panel evaluated within the pion dPDFs. We denote here $f^{\pi 
O}(x,k_\perp)=F(x,1-x,k_\perp)$ the dPDF of pion \cite{pion}. }}
\label{f1}
\end{figure}

In  our first studies,  dPDFs have been calculated within CQM, 
see Refs. \cite{noi1, noij1}~, and it has been shown that at the 
hadronic  
scale, DPC can not be neglected. For example,
as one can see in left panel of Fig. \ref{f1}, the ratio 
$uu(x_1,0.4,k_\perp)/uu(0.4,0.4,k_\perp)$ smoothly depends on $k_\perp$ 
reflecting the presence of correlations between $x_1,x_2$ and 
$k_\perp$. In fact, such a ratio would be constant if the dPDF 
factorizes as in Eq. (\ref{app}).
Here we denote $\tilde F_{uu}(x_1,x_2,k_\perp)=uu(x_1,x_2,k_\perp)$, 
being $k_\perp$ the conjugate variable to $z_\perp$ and $\tilde F$ the 
Fourier Transform of $F$.
Furthermore, in Ref. 
\cite{noij3,noice}, it has been demonstrated 
that relativistic effects, encoded in the Light-Front (LF) approach 
\cite{pol,pauli}, used here to implement the relativity in the 
calculations, introduce these kind of correlations,
 independently on the chosen model for the 
proton wave 
function nor on the energy scales dPDFs are evaluated.

In addition  in Refs. \cite{noij1,noij2}, we discuss 
the role of 
perturbative and non perturbative $x_1,x_2$ correlations in dPDFs at 
high energiy 
scales. Also in this case we found that in principle they cannot be
neglected.
In a recent work, we also show the impact of correlations in 
the pion dPDFs, see Ref. \cite{pion,pionNJL}. As one can see in the 
right panel of Fig. 
\ref{f1}, correlations are fundamental also for the pion case.
Here we  remark that within a model where the pion is described with 
two constituent quarks, the dPDF depends only on $x$ and $k_\perp$ 
since $x_2=1-x$. All these analyses show the relevance of double 
correlations in dPDFs. However, since these distributions are not 
directly accessible in experiments, in the next section we discuss the 
impact of DPC in experimental observables. 

\section{The calculation of the effective cross section}
A fundamental 
 quantity relevant for the experimental 
analyses of DPS is the so called
 effective cross section, $\sigma_{eff}$. As  previously discussed,
 this quantity represents the 
ratio between the product of one body quantities to a two body function,
 see in Eq. (\ref{sigma_eff_exp}). 
Thus, $\sigma_{eff}$ represents a valid tool to explore 
the impact of double correlations in measurements of DPS.
In Ref. \cite{noiPLB} we considered a direct evaluation of 
$\sigma_{eff}$ 
in terms of PDFs and dPDFs calculated within the same CQM.  A 
suitable expression of $\sigma_{eff}$ has been found:

\vskip -0.5cm
\begin{eqnarray}
\label{simple}
\sigma_{eff} (x_1,x_1',x_2,x_2') = 
\ddfrac{
\sum_{i,k,j,l }  F_{i} (x_{1})
F_{k} (x_{1}')
F_{j} (x_{2})
F_{l} (x_{2}')
C_{ik}
C_{jl}
}{
\sum_{i,j,k,l} 
C_{ik}
C_{jl}
\int
\tilde F_{ij}(x_1,x_2; {\bf {k_\perp}})
\tilde F_{kl}(x_1',x_2'; {\bf {-k_\perp}}) 
{d {\bf {k_\perp}} \over (2 \pi)^2}~,
}
~.
\end{eqnarray}
where here $C_{ij}$ are colour factors depending upon flavor indexes 
$i,j,k,l$, see Ref. \cite{noiPLB} for details.
 Eq. 
(\ref{simple}) reduces to a constant number if
 only one flavor and the 
full factorization ansatz (\ref{app}) would be considered. In Refs. 
\cite{noij3,noiPLB,noiads} the above quantity has 
been evaluated within different CQMs at different energy scales without 
any approximation.

\vskip 4cm

\begin{figure}[h]
\includegraphics{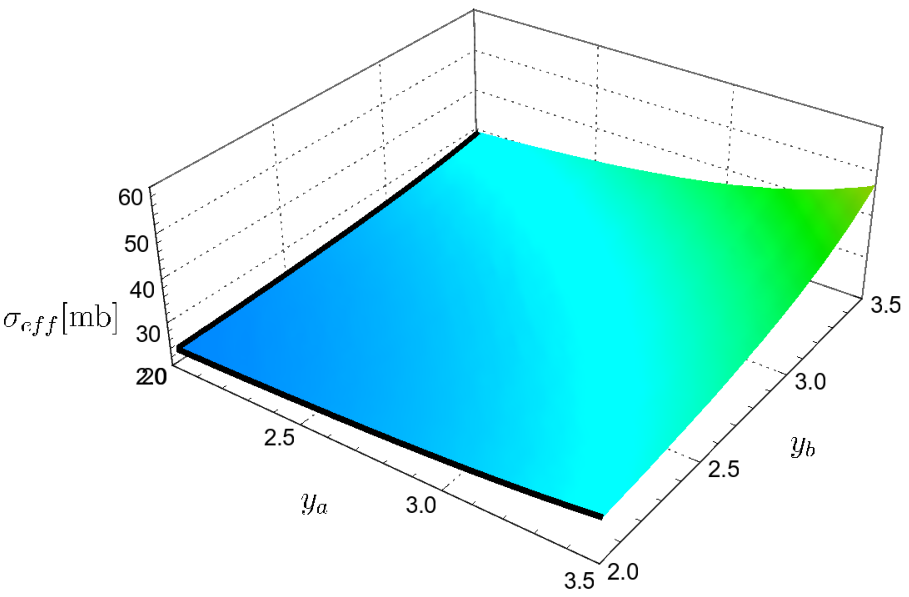}
\includegraphics{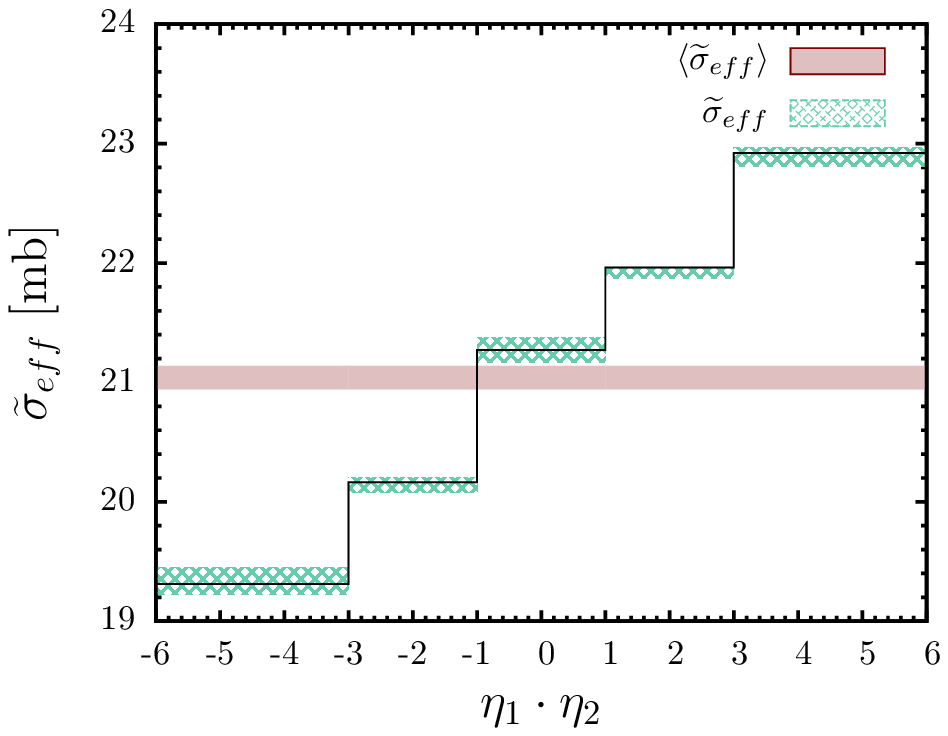}
\vskip 1.5cm
\caption{\footnotesize{ 
Left panel: $\sigma_{eff}$ Eq. (\ref{simple}) evaluated within 
relativistic QMD for gluon-gluon distributions calculated at the final 
scale $Q^2=M_W^2$ as a function of the rapidities \cite{noij3}. Right 
panel:
$\widetilde{\sigma}_{eff}$ and $\langle 
\widetilde{\sigma}_{eff} \rangle$ as a 
function of product of muon rapidities.}}
\label{efb}
\end{figure}

\noindent
As one can observe in the left panel of Fig. \ref{efb},
 once $\sigma_{eff}$ is evaluated within CQM, where correlations are 
not 
neglected, this functions strongly depends on the rapidity, i.e. on 
$x_1$ and $x_2$. These feature has been addressed in further analyses 
in Refs. \cite{noij3,noiads,pion}. In  Ref. 
\cite{noiads}, $\sigma_{eff}$ has been studied  
through dPDFs calculated within the AdS/QCD model. The mean value 
of 
$\sigma_{eff}$ and its 
 $x$ dependence, obtained within this different model,  
 is comparable with those found within the LF 
approach in Ref. \cite{noiPLB}.
Let us remark that the mean value of $\si$ evaluated in Refs. 
\cite{noiPLB,noiads} is close experimental analyses.
Let us mention that even if  the DPS 
involving mesons is still unobserved, in Ref. \cite{pion} a model 
prediction of $\si$ is provided. In this case the mean value is 
$\bar \sigma_{eff} \sim 40$ mb. Such a quantity is bigger then that 
evaluated for the proton. The interpretation of this result will be 
provided in the next section.  Let us mention that $\bar \sigma_{eff} 
$ has been used in experimental analyses, see Ref. \cite{compass}. In 
order to 
establish whether these correlations could be observed in future 
experiments, in the next section we discuss the evaluation of $\si$ for 
a specific DPS process.

\subsection{The same-sign $WW$ production at LHC}
In order to establish to which extent double correlations could be 
accessed  at the LHC, 
 in Ref. \cite{noiprl} we have studied the  
same sign $W$ pair production process, a golden channel 
 for the observation of DPS \cite{kulesza, maina, gaunt2, 
campbell}. See Ref. \cite{cotogno} for spin effects affecting this 
channel.
The differential DPS cross section can be written as follows \cite{1a}:

\vskip -0.4cm
\begin{align}
 d \sigma^{AB}_{DPS} = \dfrac{m}{2} \sum_{i,j,k,l} d \vec z_{\perp} 
F_{ij}(x_1,x_2, \vec z_\perp, \mu)F_{ij}(x_3,x_4, \vec z_\perp, \mu)
&d\hat \sigma^A_{ik} d\hat \sigma^B_{jl},
\label{dps}
\end{align}
where $\hat \sigma_{ij}^A$ represents the elementary cross section.
As non perturbative input of the calculations, use has been made of 
dPDFs evaluated in Ref. \cite{noij1} without any approximations.
 Details of the fiducial DPS phase space 
adopted in the analysis are discussed in Ref. \cite{noiprl}. In 
particular we 
found that  the total $W$-charge summed DPS cross section (considering 
both 
$W$ decays into same sign muons), calculated by means 
of dPDFs \cite{noij1}, is found to be $\sigma^{++}+\sigma^{--} 
[\mbox{fb}] \sim 
0.69$. This
result is consistent with those obtained by using for dPDFs the ansatz 
Eq. 
(\ref{app}) with PDFs evaluated with the MSTW parametrization 
\cite{MSTW} and those obtained with dPDFs of the model described in 
Ref. 
\cite{3a}.  
The effects of DPC have been investigated
by observing the
 $\tilde \sigma_{eff}$ dependence on   $\eta_1 \cdot \eta_2 \simeq 1/4~ 
\mbox{ln}(x_1/x_3)\mbox{ln}(x_2/x_4)$.  
For this process we found a mean value  $\langle \widetilde 
\sigma_{eff} \rangle \sim 21.04$ mb, consistently with calculations 
discussed in previous 
section. See Ref. \cite{noiprl} for details on the theoretical errors.
A clear signature of the presence of double 
correlations is found by observing the
departure of   $\widetilde \sigma_{eff}$ from a constant, see right 
panel of Fig. \ref{efb}.  We have  
estimated 
that, with a luminosity of $\mathcal{L} \sim 1000 
~\mbox{fb}^{-1}$, at 68\% confidence level, such a departure of 
$\sigma_{eff}$ 
from a constant value can be measured in the next run of the LHC.

\section{The link between $\sigma_{eff}$ and the transverse proton 
structure}
\begin{figure}[h]
\includegraphics{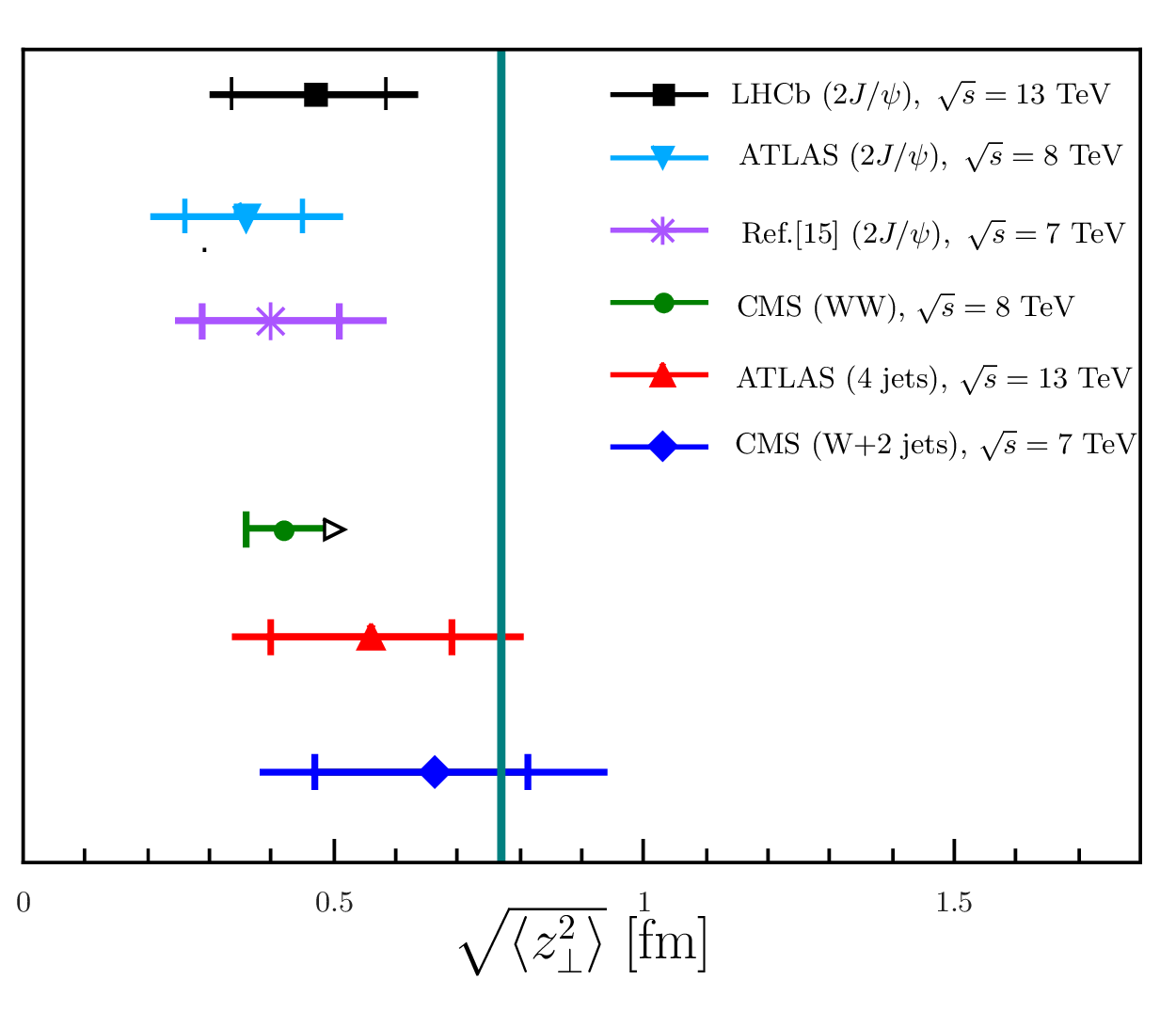}
\vskip 5.5cm
\begin{textblock*}{\textwidth}(3.5cm,-4.cm)
\tcbset{
        enhanced,
        colback=black!5!white,
        boxrule=0.1pt,
        colframe=black!75!black,
        fonttitle=\bfseries
       }
\caption{\footnotesize{ The application or Eq. (\ref{ine}) by using 
data \\ ~~~of Refs. \cite{data8,data9,data6,data10,data11,data12}.   }}
\label{f2}
\end{textblock*}
\end{figure}

While in the previous section it has been shown whether correlations 
could be observed in future experiments at the LHC, in this last part 
we show how collected experimental outcomes of $\si$ could shed some 
light on the transverse proton structure. 
To this aim here we discuss the strategy adopted in Ref. 
\cite{hadronic}. In order to interpret many of the experimental
estimates of $\si$, the factorized ansatz 
(\ref{app}) adopted in that studies has been considered.
In this scenario:

\begin{align}
 \label{si}
\si^{-1} = \int d^2 z_\perp T(z_\perp)^2 = \int \frac{d^2k_\perp}{(2 
\pi)^2} \tilde T(\vec k_\perp) \tilde T(-\vec k_\perp)~,
\end{align}
where $\tilde T(k_\perp)$ is the Fourier Transform of $T(z_\perp)$ and 
it is called effective form factor (eff).  In Refs. 
\cite{hadronic,noij3}, it 
has been shown how the probabilistic interpretation of $T(z_\perp)$ and 
the asymptotic behaviour of $\tilde T (k_\perp)$ allow to relate $\si$ 
to 
the mean transverse  distance $\langle z_\perp^2 \rangle$ 
between two active partons in a DPS process. In 
fact,  $\tilde T(0)=1$, 
$\tilde T(k_\perp \rightarrow \infty) \rightarrow 0$ and $\langle 
z_\perp^2 \rangle= -4 ~d \tilde T(k_\perp)/dk_\perp^2 ~|_{k_\perp=0} $~.
In particular one can show that in the non relativistic limit one gets:

\vskip -0.3cm
\begin{align}
 \tilde T(k_\perp) = \int dk_1~dk_1~\Psi^\dagger(\vec k_1+\vec 
k_\perp,\vec k_2)\Psi(\vec k_1, \vec k_2+\vec k_\perp)~,
\end{align}
 where $\vec k_i$ is the momentum of a parton $i$ and $\Psi$ is the 
proton 
wave function. As one can see the above expression is similar to the 
standard proton ff. However, 
since in this case there is  a double momentum 
imbalance, i.e. $k_\perp$, the eff 
can be 
considered a double form factor.
 As deeply discussed in Refs. 
\cite{hadronic,noij3}, one should expect that in the extremely high 
$k_\perp$ region, the eff should fall to zero at least as the standard 
 ff. Of course since the eff is a double ff it is reasonable that 
it goes to zero faster then the one body one, as discussed in Ref. 
\cite{hadronic,noij3,blok_1}. Thanks to these almost general and model 
independent 
conditions, we found:

\vskip -0.3cm
\begin{align}
\label{ine}
 \ddfrac{\si}{3 \pi} \leq \langle z_\perp^2 \rangle \leq 
\ddfrac{\si}{\pi}~.
\end{align}
The above expression has been validated by all models of dPDFs at 
our disposal, even for the pion target. In Fig. \ref{f2} the 
experimental values of $\si$, 
related to different processes and different final states  obtained by 
many collaborations, have been  used in Eq. (\ref{ine}) to get the mean 
transverse  distance of the active partons in a DPS process.

The above relation has been properly generalised in Ref. \cite{noij3} 
in order to include correlations between $x_1,x_2$ and $k_\perp$ going 
beyond the approximation Eq. (\ref{app}). In addition, Eq. (\ref{ine}) 
has been extended  to also include the splitting effects, i.e. the DPS 
occur 
between two partons produced by the splitting of a parent one. Within 
these results information on the transverse structure of the proton can 
be obtained from detailed experimental analyses of $\si$.

\section{Conclusions}
In this contribution we have shown and discussed results of the 
calculations of dPDFs remarking the impact of double parton correlations
and the violation of the factorzed ansatz usually assumed in 
experimental analyses.
 In 
particular we focus our attention on the role of correlations in 
experimental observables such as $\si$. In this case we found that the 
dependence of $\si$ upon the longitudinal momentum fraction of the 
partons represents a clear sign correlations. In 
particular we estimated that their effects could be observed in the 
next LHC run.
From a different perspective, our analysis also show how to unveil new 
information on the mean transverse distance of partons inside the 
proton from the present status of experimental investigations on DPS. 
All these studies point to the crucial role of dPDFs as fundamental 
tools to obtain new details on the non perturbative structure of the 
proton.

\subsection{Acknowledgments}
The author thanks all the orginzers of the conference for the support 
given for 
this talk. The author also thanks F. A. Ceccopieri, S. Scopetta, M. 
Traini and V. Vento for their collaboration to this talk. 
This  work  was  sup-ported in part by the STRONG-2020 project of the 
Euro-pean Union’s Horizon 2020 research and innovation pro-gramme under 
grant agreement No 824093.

\medskip
\section*{References}

\bibliography{iopart-num}

\providecommand{\newblock}{}
\begin{thebibliography}{10}
\expandafter\ifx\csname url\endcsname\relax
  \def\url#1{{\tt #1}}\fi
\expandafter\ifx\csname urlprefix\endcsname\relax\def\urlprefix{URL }\fi
\providecommand{\eprint}[2][]{\url{#2}}

\bibitem{hadronic}
Rinaldi M and Ceccopieri F~A 2018 {\em Phys. Rev.\/} D {\bf 97} 07150

\bibitem{noij3}
Rinaldi M and Ceccopieri F~A 2019 {\em JHEP\/} {\bf 09} 097

\bibitem{3a}
Gaunt J and Stirling W 2010 {\em JHEP\/} {\bf 03} 005

\bibitem{4a}
M~Diehl D~O and Schafer A 2012 {\em JHEP\/} {\bf 03} 089

\bibitem{5a}
Manohar A~V and Waalewijn W~J 2012 {\em Phys. Rev.\/} D {\bf 85} 114009

\bibitem{6a}
Bartalini P and Fano L (eds) 2009 {\em {Proceedings, 1st International Workshop
  on Multiple Partonic Interactions at the LHC (MPI08)}\/} DESY (Hamburg: DESY)
  ISBN 9783935702386 (\textit{Preprint} \eprint{1003.4220})

\bibitem{report}
Bansal S {\em et~al.\/} 2014 {\em {Workshop on Multi-Parton Interactions at the
  LHC (MPI @ LHC 2013) Antwerp, Belgium, December 2-6, 2013}\/}
  (\textit{Preprint} \eprint{1410.6664})

\bibitem{16a}
Aad G {\em et~al.\/} (ATLAS) 2013 {\em New J. Phys.\/} {\bf 15} 033038
  (\textit{Preprint} \eprint{1301.6872})

\bibitem{1a}
Paver N and Treleani D 1982 {\em Nuovo Cim.\/} A {\bf 70} 215

\bibitem{kase1}
Diehl M, Kasemets T and Keane S 2014 {\em JHEP\/} {\bf 05} 118
  (\textit{Preprint} \eprint{1401.1233})

\bibitem{man}
Chang H~M, Manohar A~V and Waalewijn W~J 2013 {\em Phys. Rev.\/} D {\bf 87}
  034009 (\textit{Preprint} \eprint{1211.3132})

\bibitem{noi1}
Rinaldi M, Scopetta S and Vento V 2013 {\em Phys. Rev.\/} D {\bf 87} 114021
  (\textit{Preprint} \eprint{1302.6462})

\bibitem{noij1}
Rinaldi M, Scopetta S, Traini M and Vento V 2014 {\em JHEP\/} {\bf 12} 028
  (\textit{Preprint} \eprint{1409.1500})

\bibitem{bru}
Broniowski W and Ruiz~Arriola E 2014 {\em Few Body Syst.\/} {\bf 55} 381--387
  (\textit{Preprint} \eprint{1310.8419})

\bibitem{noice}
Rinaldi M and Ceccopieri F~A 2017 {\em Phys. Rev.\/} D {\bf 95} 034040
  (\textit{Preprint} \eprint{1611.04793})

\bibitem{23a}
Kirschner R 1979 {\em Phys. Lett.\/} B {\bf 84} 266--270

\bibitem{24a}
Shelest V~P, Snigirev A~M and Zinovev G~M 1982 {\em Phys. Lett.\/} B {\bf 113}
  325

\bibitem{blok_1}
Blok B, Dokshitser {\relax Yu}, Frankfurt L and Strikman M 2012 {\em Eur. Phys.
  J.\/} C {\bf 72} 1963 (\textit{Preprint} \eprint{1106.5533})

\bibitem{noiprl}
Ceccopieri F~A, Rinaldi M and Scopetta S 2017 {\em Phys. Rev.\/} D {\bf 95}
  114030 (\textit{Preprint} \eprint{1702.05363})

\bibitem{noij2}
Rinaldi M, Scopetta S, Traini M~C and Vento V 2016 {\em JHEP\/} {\bf 10} 063
  (\textit{Preprint} \eprint{1608.02521})

\bibitem{noiPLB}
Rinaldi M, Scopetta S, Traini M and Vento V 2016 {\em Phys. Lett.\/} B {\bf
  752} 40--45 (\textit{Preprint} \eprint{1506.05742})

\bibitem{noiads}
Traini M, Rinaldi M, Scopetta S and Vento V 2017 {\em Phys. Lett.\/} B {\bf
  768} 270--273 (\textit{Preprint} \eprint{1609.07242})

\bibitem{MPI15}
Jung H, Treleani D, Strikman M and van Buuren N (eds) 2016 {\em {Proceedings,
  7th International Workshop on Multiple Partonic Interactions at the LHC
  (MPI@LHC 2015)}\/} ISBN 9783945931011
  \urlprefix\url{https://bib-pubdb1.desy.de/record/297386}

\bibitem{S1}
Åkesson T {\em et~al.\/} (Axial Field Spectrometer) 1987 {\em Z. Phys.\/} C
  {\bf 34} 163

\bibitem{S2}
Alitti J {\em et~al.\/} (UA2) 1991 {\em Phys. Lett.\/} B {\bf 268} 145--154

\bibitem{S3}
Abe F {\em et~al.\/} (CDF) 1997 {\em Phys. Rev.\/} D {\bf 56} 3811--3832

\bibitem{S4}
Abazov V~M {\em et~al.\/} (D0) 2010 {\em Phys. Rev.\/} D {\bf 81} 052012
  (\textit{Preprint} \eprint{0912.5104})

\bibitem{S5}
Aad G {\em et~al.\/} (ATLAS) 2013 {\em New J. Phys.\/} {\bf 15} 033038
  (\textit{Preprint} \eprint{1301.6872})

\bibitem{S6}
Chatrchyan S {\em et~al.\/} (CMS) 2014 {\em JHEP\/} {\bf 03} 032
  (\textit{Preprint} \eprint{1312.5729})

\bibitem{S7}
Aaij R {\em et~al.\/} (LHCb) 2014 {\em JHEP\/} {\bf 04} 091 (\textit{Preprint}
  \eprint{1401.3245})

\bibitem{data8}
Aaij R {\em et~al.\/} (LHCb) 2017 {\em JHEP\/} {\bf 06} 047 [Erratum:
  JHEP10,068(2017)] (\textit{Preprint} \eprint{1612.07451})

\bibitem{data9}
Chatrchyan S {\em et~al.\/} (CMS) 2014 {\em JHEP\/} {\bf 03} 032
  (\textit{Preprint} \eprint{1312.5729})

\bibitem{data6}
Aaboud M {\em et~al.\/} (ATLAS) 2016 {\em JHEP\/} {\bf 11} 110
  (\textit{Preprint} \eprint{1608.01857})

\bibitem{data10}
Sirunyan A~M {\em et~al.\/} (CMS) 2018 {\em JHEP\/} {\bf 02} 032
  (\textit{Preprint} \eprint{1712.02280})

\bibitem{data11}
Aaboud M {\em et~al.\/} (ATLAS) 2017 {\em Eur. Phys. J.\/} C {\bf 77} 76
  (\textit{Preprint} \eprint{1612.02950})

\bibitem{data12}
Lansberg J~P and Shao H~S 2015 {\em Phys. Lett.\/} B {\bf 751} 479--486
  (\textit{Preprint} \eprint{1410.8822})

\bibitem{pol}
Keister B~D and Polyzou W~N 1991 {\em Adv. Nucl. Phys.\/} {\bf 20} 225--479
  [,225(1991)]

\bibitem{pauli}
Brodsky S~J, Pauli H~C and Pinsky S~S 1998 {\em Phys. Rept.\/} {\bf 301}
  299--486 (\textit{Preprint} \eprint{hep-ph/9705477})

\bibitem{pion}
Rinaldi M, Scopetta S, Traini M and Vento V 2018 {\em Eur. Phys. J.\/} C {\bf
  78} 781 (\textit{Preprint} \eprint{1806.10112})

\bibitem{pionNJL}
Courtoy A, Noguera S and Scopetta S 2019  (\textit{Preprint}
  \eprint{1909.09530})

\bibitem{compass}
Koshkarev S 2019 {\em {18th Workshop on High Energy Spin Physics (DSPIN-19)
  Dubna, Joint Institute for Nuclear Research (JINR), Moscow region, Russia,
  September 2-6, 2019}\/} (\textit{Preprint} \eprint{1909.06195})

\bibitem{kulesza}
Kulesza A and Stirling W~J 2000 {\em Phys. Lett.\/} B {\bf 475} 168--175
  (\textit{Preprint} \eprint{hep-ph/9912232})

\bibitem{maina}
Maina E 2009 {\em JHEP\/} {\bf 09} 081 (\textit{Preprint} \eprint{0909.1586})

\bibitem{gaunt2}
Gaunt J~R, Kom C~H, Kulesza A and Stirling W~J 2010 {\em Eur. Phys. J.\/} C
  {\bf 69} 53--65 (\textit{Preprint} \eprint{1003.3953})

\bibitem{campbell}
Campbell J~M, Ellis R~K and Williams C 2011 {\em JHEP\/} {\bf 07} 018
  (\textit{Preprint} \eprint{1105.0020})

\bibitem{cotogno}
Cotogno S, Kasemets T and Myska M 2019 {\em Phys. Rev.\/} D {\bf 100} 011503
  (\textit{Preprint} \eprint{1809.09024})

\bibitem{MSTW}
Martin A~D, Stirling W~J, Thorne R~S and Watt G 2009 {\em Eur. Phys. J.\/} C
  {\bf 63} 189--285 (\textit{Preprint} \eprint{0901.0002})

\end{thebibliography}


\begin{thebibliography}{9}





\bibitem{hadronic}	
  M.~Rinaldi and F.~A.~Ceccopieri,
  Phys.\ Rev.\ D {\bf 97}, no. 7, 071501 (2018)

\bibitem{noij3}	
  M.~Rinaldi and F.~A.~Ceccopieri,
  JHEP {\bf 1909}, 097 (2019)








\bibitem{3a}  
  J.R. Gaunt and W.J. Stirling,
  JHEP {\bf 03}, 005 (2010)


\bibitem{4a}  
  M. Diehl, D. Ostermeier and A. Schafer,
JHEP {\bf 03}, 089 (2012)



\bibitem{5a} 
  A. V. Manohar and  W. J. Waalewijn,
  Phys.\ Rev.\ D {\bf 85}, 114009 (2012)

\bibitem{6a} 
 P. Bartalini (ed.)  and L. Fan\`o (ed.), 
arXiv:1003.4220 [hep-ex]



\bibitem{report} 
 S.~Bansal  {\it et al.},
  arXiv:1410.6664 [hep-ph].

  
  \bibitem{16a} 
 G. Aad, { et al.}  [ATLAS Collaboration],
New J.\ Phys.\  {\bf 15}, 033038 (2013) 
  
  
  
  
  
  \bibitem{1a} 
 N. Paver and D. Treleani,
Nuovo Cim.\ A {\bf 70}  215 (1982) 


  
  \bibitem{kase1}  
  M.~Diehl, T.~Kasemets and S.~Keane,
evolution,''
  JHEP {\bf 05}, 118 (2014)
  
  


  
  \bibitem{man}  
 H.M.  Chang, A.V. Manohar and W.J. Waalewijn,
  Phys.\  Rev.\  D 87, {\bf 034009} (2013) 


  \bibitem{noi1} 
 M. Rinaldi, S. Scopetta and V. Vento,
  Phys.\ Rev.\ D {\bf 87}, 11,  114021 (2013)  
  
  
\bibitem{noij1}
  M.~Rinaldi, S.~Scopetta, M.~Traini and V.~Vento,
Front 
  JHEP {\bf 12}, 028 (2014).
  
  
\bibitem{bru}
  W.~Broniowski and E.~Ruiz Arriola,
model,''
  Few Body Syst.\  {\bf 55}, 381 (2014).
  
  
  
    \bibitem{noice} 
  M.~Rinaldi and F.~A.~Ceccopieri,
  Phys.\ Rev.\ D {\bf 95}, no. 3, 034040 (2017).
  
  

  
  
  
  
  
  \bibitem{23a} 
 R. Kirschner,
  Phys.\ Lett.\ B {\bf 84}, 266 (1979)


\bibitem{24a} 
 V.P. Shelest, A.M. Snigirev and G.M. Zinovev,
  Phys.\ Lett.\ B {\bf 113}, 325 (1982)
  
    
\bibitem{blok_1}  
  B.~Blok {\it et al.},
  Eur.\ Phys.\ J.\ C {\bf 72}, 1963 (2012);
  Eur.\ Phys.\ J.\ C {\bf 74}, 2926 (2014).



 \bibitem{noij2}    
  M.~Rinaldi, S.~Scopetta, M.~C.~Traini and V.~Vento,
  JHEP {\bf 10}, 063 (2016).

  
  
         \bibitem{noiPLB}  
  M.~Rinaldi, S.~Scopetta, M.~Traini and V.~Vento,
within a 
  Phys.\ Lett.\ B {\bf 752}, 40 (2016).
  

\bibitem{noiads}  
  M.~Traini, M.~Rinaldi, S.~Scopetta and V.~Vento,
  Phys.\ Lett.\ B {\bf 768}, 270 (2017).
  
  
  
  
  
  
  
  
  
\bibitem{MPI15}   
H.~Jung \textsl{\& al.}, Proceedings, 7th International Workshop on 
Multiple
                        Partonic Interactions at the LHC (MPI@LHC 2015),
https://bib-pubdb1.desy.de/record/297386, DESY-PROC-2016-01


\bibitem{S1}  
T. Akesson et al. [Axial Field Spectrometer Collaboration], 
Z. 
Phys. C {\bf 34} (1987) 163.

\bibitem{S2} 
J. Alitti et al. [UA2 Collaboration], Phys. Lett. {\bf 
B268}, 145 
(1991).

\bibitem{S3} 
F. Abe et al. [CDF Collaboration], Phys. Rev. D {\bf 56}, 
3811 
(1997).

\bibitem{S4} 
V. M. Abazov et al. [D0 Collaboration], Phys. Rev. D {\bf 
81}, 
052012 (2010).

\bibitem{S5} 
G. Aad et al. [ATLAS Collaboration], New J. Phys. 15, 
033038 
(2013).

\bibitem{S6}  
S. Chatrchyan et al. [CMS Collaboration], JHEP 03, 032 
(2014).

\bibitem{S7}
  R.~Aaij {\it et al.} [LHCb Collaboration],
  JHEP {\bf 04} (2014) 091

  


  
  
   \bibitem{noiprl} 
  F.~A.~Ceccopieri, M.~Rinaldi and S.~Scopetta,
  arXiv:1702.05363 [hep-ph].

  \bibitem{pol} 
   B.D. Keister and W.N. Polyzou,
  Adv.\ Nucl.\ Phys.\  {\bf 20}, 225 (1991) 

\bibitem{pauli} 
  S.J. Brodsky, H.C. Pauli and S.S. Pinsky,
  Phys.\ Rept.\  {\bf 301}, 299 (1998) 
  
 

  
  \bibitem{faccioli} 
 P. Faccioli, M. Traini and V. Vento,
  Nucl.\ Phys.\ A {\bf 656}, 400 (1999) 
  
  
  
  
    \bibitem{MSTW} 
  A.~D.~Martin, W.~J.~Stirling, R.~S.~Thorne and G.~Watt,
  Eur.\ Phys.\ J.\ C {\bf 63}, 189 (2009).
  
  
  

  
\bibitem{afs} 
  T.~Akesson {\it et al.} [Axial Field Spectrometer Collaboration],
$\sqrt{s}=63$-{GeV},''
  Z.\ Phys.\ C {\bf 34}, 163 (1987).

\bibitem{data0} 
  J.~Alitti {\it et al.} [UA2 Collaboration],
search 
  Phys.\ Lett.\ B {\bf 268}, 145 (1991).


\bibitem{data2} 
  F.~Abe {\it et al.} [CDF Collaboration],
  Phys.\ Rev.\ D {\bf 56}, 3811 (1997).
  
  
  
   
  
  
  
  

  
\bibitem{kulesza}
  A.~Kulesza and W.~J.~Stirling,
  Phys.\ Lett.\ B {\bf 475}, 168 (2000).


\bibitem{maina}
  E.~Maina,
  JHEP {\bf 09}, 081 (2009).


\bibitem{gaunt2}
  J.~R.~Gaunt \textit{et a.}, 
  Eur.\ Phys.\ J.\ C {\bf 69}, 53 (2010).





\bibitem{campbell} 
  J.~M.~Campbell, R.~K.~Ellis and C.~Williams,
  JHEP {\bf 07}, 018 (2011).
  
  
  
  
  
  
  
  

  
  \bibitem{catani} 
  S.~Catani \textsl{et al.}, 
  \textsl{Phys.~Rev.~Lett.~} {\bf 103}, 082001 (2009);
  S.~Catani and M.~Grazzini,
  \textsl{Phys.~Rev.~Lett.~} {\bf 98}, 222002 (2007).
  




\bibitem{pion} 
  M.~Rinaldi, S.~Scopetta, M.~Traini and V.~Vento,
  Eur.\ Phys.\ J.\ C {\bf 78}, no. 9, 781 (2018)



\bibitem{pionNJL} 
  A.~Courtoy, S.~Noguera and S.~Scopetta,
  arXiv:1909.09530 [hep-ph].
  
 
\bibitem{compass} 
  S.~Koshkarev,
  arXiv:1909.06195 [hep-ph].
  
 
\bibitem{cotogno} 
  S.~Cotogno, T.~Kasemets and M.~Myska,
  Phys.\ Rev.\ D {\bf 100}, no. 1, 011503 (2019)






\bibitem{data8} 
  R.~Aaij {\it et al.} [LHCb Collaboration],
  JHEP {\bf 06} (2017) 047
   Erratum: [JHEP {\bf 10} (2017) 068].



  \bibitem{data6} 
  M.~Aaboud {\it et al.} [ATLAS Collaboration],
  JHEP {\bf 11}, 110 (2016).


 
 


 
 \bibitem{data9} 
  S.~Chatrchyan {\it et al.} [CMS Collaboration],
  JHEP {\bf 03}, 032 (2014)
 
 
 \bibitem{data10}
  A.~M.~Sirunyan {\it et al.} [CMS Collaboration],
  JHEP {\bf 1802}, 032 (2018)
  
  
  
  \bibitem{data11} 
  M.~Aaboud {\it et al.} [ATLAS Collaboration],
  Eur.\ Phys.\ J.\ C {\bf 77}, no. 2, 76 (2017)
  
  
  \bibitem{data12}
  J.~P.~Lansberg and H.~S.~Shao,
  Phys.\ Lett.\ B {\bf 751}, 479 (2015)
 





  
  
\end{thebibliography}

\smallskip

\end{document}